\documentstyle[NATO,numreferences,epsf]{CRCKAPB}

\begin{opening}
\title{WHITE DWARF COLORS IN LOW ACCRETION RATE
BINARIES}

\author{DEAN M. TOWNSLEY}
\institute{Department of Physics, University of California, Santa Barbara
Broida Hall, Santa Barbara, CA 93106}
\author{LARS BILDSTEN}
\institute{Kavli Institute for Theoretical Physics and Department of Physics,
University of California, Santa Barbara\\Kohn Hall, Santa Barbara, CA 93106}

\end{opening}

\begin{document}

\begin{abstract}

Our recent theoretical work (Townsley and Bildsten 2002) on the thermal
state of white dwarfs (WDs) in low mass transfer rate binaries allows us to
predict the broadband colors of the binary from those of the WD and companion
when the disk is dim. The results based on standard CV evolution are presented here.
These will aid the discovery
of such objects in field surveys and proper-motion selected globular cluster
surveys with \emph{HST}; especially for the largely unexplored post period minimum
Cataclysmic Variables (CVs) with
the lowest accretion rates and degenerate companions. We have also calculated
the fraction of time that the WD resides in the ZZ Ceti instability strip
thus clarifying that we expect many accreting WDs to exhibit non-radial
oscillations.  The study of these will provide new insights into the
rotational and thermal structure of an actively accreting WD.

\end{abstract}

\section{Introduction}
As summarized elsewhere in these proceedings our recent work has
demonstrated that at the low accretion rates appropriate for dwarf novae (DN), the core
temperature, $T_c$, of the WD is \emph{set} by the long-time average accretion rate,
$\langle \dot M\rangle$, of the binary \cite{TownBild02}.  With $T_c$ set,
we determine the WD luminosity, which depends on $\langle \dot
M\rangle$, $M_{\rm WD}$, and the mass of the freshly accreted layer.  This luminosity
is directly observable for DN when the disk is in quiescence and the surface of the WD
has cooled after outburst.

\section{Results}
\begin{figure}
\vspace{-0.1cm}
\epsfysize=7.3cm \epsfbox{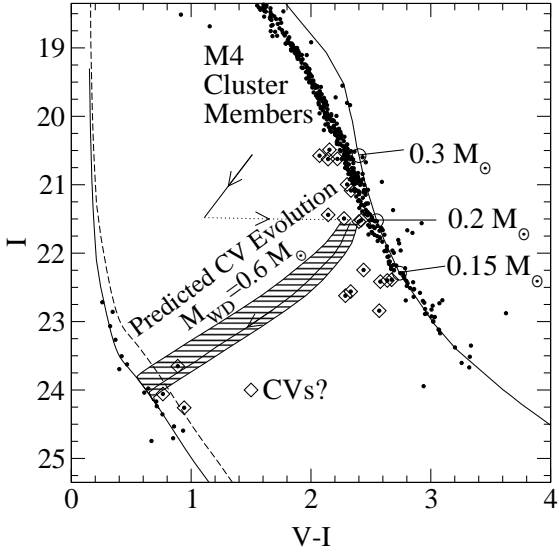}
\vspace{-0.2cm}
\caption{
\label{fig:M4}
Evolutionary path for a $0.6M_\odot$ WD in a CV.  For simplicity,
only contributions from the WD and MS companion have been included.
The displayed \emph{HST} data are proper motion selected cluster
members [4].
}
\vspace{-0.1cm}
\end{figure}

In the quiescent state, the broadband colors are just that of a hot WD plus a low
mass main sequence star or brown dwarf companion.  Figure \ref{fig:M4} shows the color
and magnitude we predict for quiescent CV binaries
using standard CV evolution to relate the companion mass to $\langle\dot M\rangle$
\cite{Howeetal01,KoldBara99}.  The data shown for comparison are members of the globular
cluster M4 as determined by proper motion studies with \emph{HST} \cite{King98}.
The $0.6M_\odot$ WD primary
spends approximately 1 Gyr in the ZZ Ceti instability strip.

This work was supported by the NSF under Grants PHY99-07949,
AST01-96422, and AST02-05956, and by
NASA through grant AR-09517.01-A from STScI, which is operated by
AURA, Inc., under NASA contract NAS5-26555. 
L. B. is a Cottrell Scholar of the Research Corporation and D. T. is an NSF Graduate
Fellow.

\end{document}